\documentclass[apl,twocolumn,reprint,floatfix,superscriptaddress]{revtex4}

\usepackage{graphicx}
\usepackage{dcolumn}   % needed for some tables
\usepackage{bm}        % for math
\usepackage{amssymb}   % for math
\usepackage{verbatim}
\usepackage{multirow}

\usepackage{amsmath}
\usepackage{amsfonts}

\begin{document}

\title{Superconducting microstrip amplifiers with sub-Kelvin noise temperature near 4 GHz}

\author{M.P. DeFeo}
\affiliation{Department of Physics, Syracuse University, Syracuse, NY 13244-1130}
\author{B.L.T. Plourde}
\email[]{bplourde@phy.syr.edu}
\affiliation{Department of Physics, Syracuse University, Syracuse, NY 13244-1130}

\date{\today}

\begin{abstract}
We present measurements of an amplifier operating at 3.8 GHz with 150 MHz of bandwidth based on the microstrip input-coil resonance of a dc superconducting quantum interference device (SQUID) with submicron Josephson junctions. The noise temperature is measured using two methods: comparing the signal-to-noise ratio of the system with and without the SQUID in the amplifier chain, and using a modified Y-factor technique where calibrated narrowband noise is mixed up to the SQUID amplifier operating frequency. With the SQUID cooled to 0.35 K we observe a minimum system noise temperature of 0.55 $\pm~0.13$ K, dominated by the contribution from the SQUID amplifier.
\end{abstract}

\maketitle
Superconducting circuits coupled to microwave cavities have provided a valuable architecture for studying quantum mechanics using macroscopic systems where electronic circuit degrees of freedom are treated as quantum variables \cite{2004Natur.431..162W}. The small signals characteristic of these systems have driven substantial interest in amplifiers with near quantum-limited noise performance that operate in the microwave frequency range. There have been many recent advances in Josephson-based amplifiers including superconducting undulatory galvanometers (SLUGs) \cite{2012ApPhL.100f3503H} and Josephson parametric amplifiers \cite{2010Natur.465...64B}. Microstrip superconducting quantum interference device amplifiers \cite{2010SuScT..23i3001M} (MSAs) based on resonant microstrip input coils have recently demonstrated their utility in qubit measurement --- an MSA was used to perform dispersive readout of a flux qubit \cite{2011PhRvB..84v0503J}. A scheme to perform a quantum non-demolition like measurement of a flux qubit using an MSA has also been proposed \cite{2008PhRvB..78e4507S}. 

Microstrip SQUID amplifiers employ a dc SQUID washer inductively coupled to a $\lambda /2$ resonant transmission line that serves as the device input coil \cite{1998ApPhL..72.2885M}. Signals are driven between one end of the coil and ground, where the other end of the coil is left unterminated and the SQUID washer is typically grounded. The stripline mode between the input coil and the washer allows oscillatory signals to propagate with a resonant enhancement of the current flowing through the coil when the signal frequency is near the standing-wave resonance set by the length of the coil. MSAs operated at low temperatures are capable of achieving low noise temperatures and have demonstrated near quantum-limited noise performance at 500 MHz \cite{2001ApPhL..78..967M}. More recently, an alternative configuration with a small-area SQUID coupled to a lumped-element quarter-wave resonator was shown to operate as an amplifier in the gigahertz range \cite{2008ApPhL..93h2506S}.

The operating frequency of an MSA is determined by the electrical length of the input coil. Thus, to make these devices operate at higher frequency the input coil must be shortened. However, this reduces the mutual inductance and hence reduces the gain of the amplifier \cite{2003ApPhL..82.3266M}. In a previous letter \cite{2010ApPhL..97i2507D} we reported a scheme to enhance the gain of MSAs by replacing the large area junctions common to most microstrip SQUID amplifiers with submicron shadow-evaporated $\mathrm{Al-AlO_x-Al}$ Josephson junctions. By reducing the junction area, the capacitance $C$ associated with that area is also reduced allowing the use of large shunt resistors $R$ while keeping the SQUID response single valued. The gain $G$ of an MSA is proportional to $M_i^2 V_{\Phi}^2$, where $M_i$ is the mutual inductance between the input coil and the SQUID washer and $V_{\Phi}$ is the flux-to-voltage transfer coefficient, which can be approximated as $R/L_{SQ}$, where $L_{SQ}$ is the SQUID self-inductance. Thus, larger shunt resistors increase $V_{\Phi}$ while maintaining a nonhysteretic device ($\beta_c \equiv 2 \pi I_0 R^2 C / \Phi_0 < 1$, where $I_0$ is the junction critical current and $\Phi_0 \equiv h/2e$). This enhancement of $V_\Phi$ offsets the reduction in $M_i$ from shortening the input coil. In this letter, we report an alternative design of the MSA input coil and washer. This design, combined with our technique to boost the gain, results in an MSA operating at 3.8 GHz with 17 dB of gain and a 150 MHz bandwidth. In addition, we characterize the system noise temperature and demonstrate a substantial reduction compared to a conventional cryogenic HEMT amplifier alone. 

\begin{figure}[hb]
\centering
\includegraphics[width=3.35in]{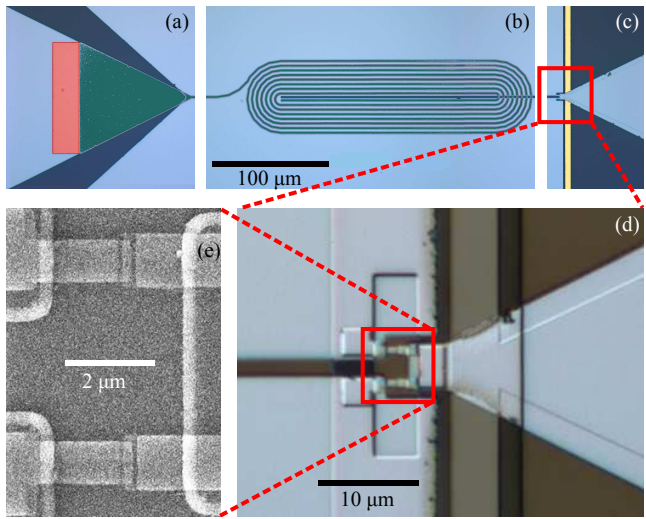}
  \caption{(Color online) (a) False-color optical micrograph of coupling capacitor (red), (b) input coil (green) and (c) Pd shunt resistors (yellow) at scale specified in (b). (d) Closeup optical micrograph of junction and shunt region. (e) Scanning electron micrograph of Josephson junctions.
\label{fig:setup}}
\end{figure}

Devices are fabricated at the 100 mm wafer level on thermally oxidized silicon wafers. The SQUID loop is formed by a large Al washer that also serves as a ground plane for the microstrip input coil (Fig. 1). The SQUID inductance, estimated to be $L_{SQ} = 90~\mathrm{pH}$, is formed by a $2~\mathrm{\mu m}$ x $310~\mathrm{\mu m}$ slit in the SQUID washer. The input coil is formed by depositing Al into a lift-off photoresist mask on top of $150~\mathrm{nm}$ of PECVD deposited $\mathrm{SiO_2}$ over the ground plane, which provides dielectric insulation between the input coil and ground. The length of the input coil is defined by an on-chip coupling capacitor estimated to be $C_c \approx 0.3~\mathrm{pF}$ coupling the coil to the input contact pad. For a small inductance washer with only a slit to form the SQUID loop, we employ a `racetrack' input coil geometry to provide an enhanced mutual inductance between the input coil and SQUID washer for a given coil length when compared with a more traditional spiral input coil. The mutual inductance between the input coil and SQUID washer is $M_i \approx n\alpha L_{SQ}$ where $n$ is the number of turns of the coil and $\alpha$ is the fraction of the total slit length enclosed by the center turn of the input coil \cite{1982ApPhL..40..736K}. For a given coil length, the racetrack geometry leads to a greater $\alpha$ when compared to a more conventional spiral input coil. Although the number of turns is less than in the spiral geometry, the larger $\alpha$ results in an enhancement of mutual inductance $M_i \approx 0.5~\mathrm{nH}$ for a $2~\mathrm{\mu m}$ linewidth coil and $2~\mathrm{\mu m}$ spacing between each of its 8 turns.

With the exception of the junction electrodes, all layers are patterned photolithograpically. The tunnel junctions are patterned with electron-beam lithography and formed using a standard double-angle shadow evaporation technique \cite{1977ApPhL..31..337D}. Based on various process calibrations and room-temperature characterization of the device, we estimate a critical current per junction of $3.5~\mu\mathrm{A}$ and a resistance per shunt of $63~\Omega$.

The device was mounted to a custom microwave board with multiple short Al wirebonds connecting the input/output pads to copper traces on the board that interface with coaxial cables through microwave connectors to carry signals to and from the device. The SQUID washer was wirebonded to the ground of the microwave board around its perimeter to provide a low-impedance path to ground. The device was enclosed in an Al box that contained a wire-wound coil for flux biasing the SQUID. The SQUID was current biased through a microwave bias-T separated from the output of the device by a 2 dB attenuator for impedance matching. The bias current and flux bias for the SQUID were supplied with batteries and the lines were passed through cryogenic Cu-powder filters. 

Cryogenic measurements were performed on a Janis $^3\mathrm{He}$ refrigerator at a base temperature of $0.35~\mathrm{K}$. Device gain was determined by measuring the transmission, $\left|S_{21}\right|$, using a vector network analyzer to supply a weak drive ($-125~\mathrm{dBm}$ at the MSA input). The drive signal was sent down a coaxial cable with 10 dB of attenuation at 4 K and 30 dB at 0.35 K at the input of the MSA. The signal from the MSA was carried to the first stage of amplification, a CITCRYO1-12A HEMT amplifier mounted on the 4 K plate of the refrigerator, through a superconducting niobium coaxial cable. A room-temperature microwave amplifier was used to boost the signal further. Gain measurements were performed at the optimum flux and current bias points and a maximum gain of 17 dB was observed at a frequency of 3.8 GHz (Fig. 2) relative to a system baseline measured on a separate cooldown. The structure in our measurement of gain versus frequency (Fig. 2) indicates that there is likely a moderate impedance mismatch between our MSA and the signal cables, particularly for frequencies below the input coil resonance. The origin of this frequency dependence of the mismatch is not known. We note that a detailed characterization of the MSA input impedance could be performed in future experiments with a different measurement setup following techniques similar to those described in Refs. \cite{2008ApPhL..92q2503K,2008ApPhL..93h2506S}. However, such a characterization is beyond the scope of our present work. The gain was measured as a function of drive power and the 1 dB compression point of the device was observed to be $-117~\mathrm{dBm}$ at the input of the MSA [Fig. 2 (inset)]. The small initial increase in gain at low input powers may be related to the presence of resonant structure on the $V_\Phi$ for MSAs with tightly coupled coils \cite{2010ApPhL..97i2507D} where a particular input signal strength could optimize the sampling of steep regions of the $V_\Phi$ response.

\begin{figure}[hb]
\centering
\includegraphics[width=3.35in]{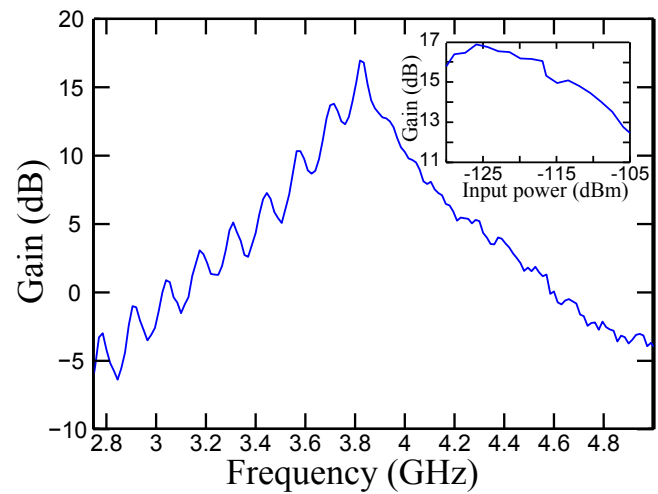}
  \caption{(Color online) Device gain as a function of frequency biased for maximum gain. (Inset) Maximum device gain as a function of drive power.
\label{fig:setup}}
\end{figure}

Noise properties of the amplifier chain were measured using two methods: comparing the signal-to-noise ratio (SNR) with and without the MSA in the measurement chain, and using a modified narrow-band Y-factor measurement technique. For a given single-tone microwave drive incident on the measurement chain, the SNR was measured at the output of the system. This measurement was performed in two cooldowns, one with the MSA in the measurement circuit, and one without the MSA. This SNR-ratio technique is sensitive to any impedance mismatches between the MSA and the input and output circuitry, therefore it provides an upper bound on the noise temperature of our system. Comparing the SNR of these two measurements at 4 GHz yields a maximum SNR increase of 7.2 dB with the MSA in the circuit. For our measured HEMT noise temperature of 3.1 K, that corresponds to a system noise temperature of 0.59 K with the MSA in the measurement chain. We measured the SNR with the MSA included in the circuit over the full frequency span in Fig. 4(b), however the HEMT SNR was only measured at 4.0 GHz. Thus, for our analysis we assumed the gain and noise of the HEMT to be constant over the frequency span of Fig. 4(b). The error bars on the SNR-ratio points were estimated from the small expected variation in these quantities based on the HEMT data sheet.

\begin{figure}[hb]
\centering
\includegraphics[width=3.35in]{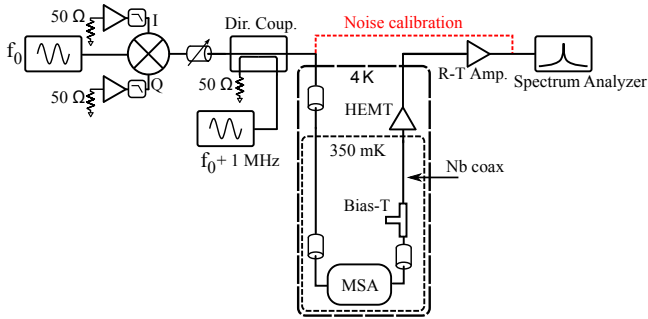}
  \caption{(Color online) Schematic of modified narrowband Y-factor noise temperature measurement circuit. Dashed red line indicates signal path for calibration measurement.
\label{fig:setup}}
\end{figure}

We performed an alternative investigation of the system noise using a modified narrowband Y-factor measurement. Instead of using the Johnson-Nyquist noise of a temperature-controlled resistor as a signal source as in a conventional Y-factor measurement, noise from room-temperature amplifiers was used to drive the device. Two SR560 amplifiers were terminated at their input with broadband $50~\Omega$ loads and were low-pass filtered at 300 kHz. These noise signals were mixed up to the operating frequency of the MSA with an I-Q mixer where the local oscillator was provided by a microwave generator and each noise source was connected to the I or Q port. Simply injecting amplified thermal noise across the MSA bandwidth would exceed the dynamic range of the MSA for relatively weak noise powers. Instead, by supplying noise over only a narrow  bandwidth centered at a frequency of interest in the MSA operating range, it is possible to vary the injected noise power over a much larger range, making an extraction of the amplifier noise feasible. 

The injected noise power was controlled with a variable attenuator between the signal source and the input at the top of the refrigerator. For each setting of the attenuator, the narrow-band noise signal was calibrated at the output of the directional coupler by directly measuring the power spectrum of the noise using a spectrum analyzer. Taking into account the loss between the point of calibration and the input of the MSA, in a given resolution bandwidth the measured noise power is related to the Johnson-Nyquist noise of a matched resistive load at a temperature $T_{eff}$ at the input of the MSA. Varying the noise power at the input of the MSA is analogous to changing the temperature of a resistive load, the technique common to traditional Y-factor measurements, but difficult to implement for a cryogenic amplifier. A  second microwave generator was used to produce a calibration tone, displaced from the center frequency of the noise signal by 1 MHz, and this was combined with the noise signal through a directional coupler (Fig. 3). The peak height of this tone was used to monitor the gain of the device during the measurement. Although this technique presents many advantages to a traditional Y-factor measurement, it depends on the accuracy of the calibration of the total loss of the drive line. We estimate a systematic uncertainty of $\pm~ 1~\mathrm{dB}$ on this calibration based on a variety of room-temperature and cryogenic measurements of individual microwave components. This systematic uncertainty is accounted for with error bars with upper and lower limits set by the calculated noise temperature for a given input $T_{eff}$ with $\pm~ 1~\mathrm{dB}$ of extra loss on the drive line. 

\begin{figure}[hb]
\centering
\includegraphics[width=3.35in]{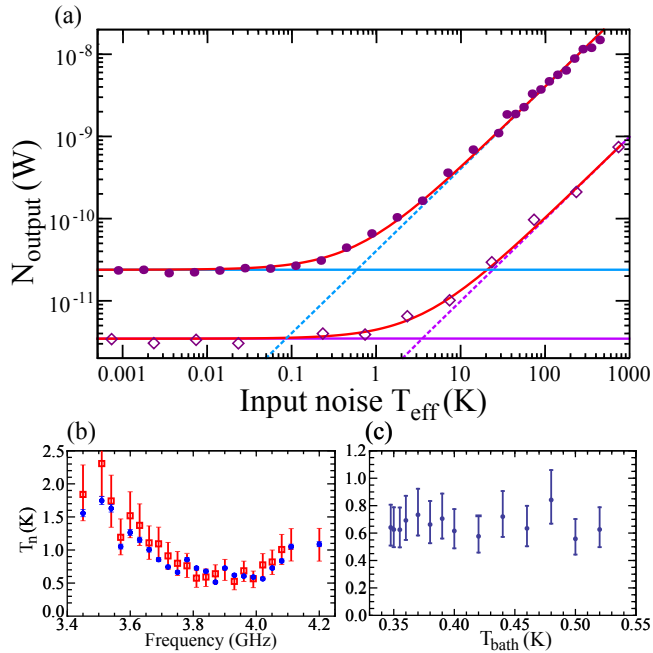}
  \caption{(Color online) (a) Power measured at the output of the measurement chain as a function of input noise $T_{eff}$ for the HEMT amplifier alone (open diamonds) and with the MSA included (circles). The data is fit to a line (red) and the system noise temperature corresponds to the intersection of the fit offset (horizontal line) and slope (dashed line). (b) System noise temperature at bath temperature of $0.35 \mathrm{K}$ as a function of frequency. Blue circles correspond to SNR-ratio measurements and open red squares were measured using the modified Y-factor technique. (c) System noise temperature at a frequency of 3.83 GHz as a function of bath temperature measured using the modified Y-factor technique.
\label{fig:setup}}
\end{figure}

For a given center frequency of the input noise, the $T_{eff}$ corresponding to several noise powers was calculated. This noise was then used to drive the amplifier chain and the system output noise power was measured. This modified Y-factor measurement technique was implemented to perform a noise temperature measurement of a known quantity, our HEMT amplifier (with the MSA removed from the circuit), as a proof-of-principle demonstration. The acquired data is plotted on a log-log scale and the offset and slope of the data are fit and plotted independently, and the system noise temperature occurs at the intersection of these two lines. The offset and slope of the data are combined and plotted as a single line to fit the data.  The result of this measurement [Fig. 4(a)] yields a system noise temperature of 3.1 K, which was in agreement with the HEMT data sheet provided by the manufacturer. In a subsequent cooldown the system noise temperature was again measured, but now with the MSA in the measurement circuit. With the MSA biased for maximum gain, a minimum system noise temperature of 0.55 K $\pm~0.13~\mathrm{K}$ was observed at a frequency of 3.8 GHz [Fig. 4(a)]. For the same bias conditions used for the noise temperature measurement at 3.8 GHz, the system noise temperature was measured as a function of frequency indicating the minimum system noise occurred near the maximum gain of the MSA [Fig 4(b)]. 

The noise properties of SQUID amplifiers have been studied extensively with numerical simulations \cite{1983ApPhL..43..694H,1977JLTP...29..301T,1979JLTP...37..405C,1985JLTP...61..227M}. For sufficiently high frequencies, the noise floor of a SQUID amplifier is dominated by Johnson-Nyquist noise of the resistive shunts, and thus scales with the electron temperature in the shunts. For a tuned SQUID amplifier operating at $\omega_0$ with optimal noise matching to the source impedance, the noise temperature is expected to scale as $T_N^{opt}  \propto (\omega_0/V_\Phi) T$. We studied the temperature dependence of the noise of our MSA at 3.85 GHz by varying the bath temperature over nearly a 200 mK range [Fig. 4(c)]. No significant variation of $T_N$ was observed, suggesting that the electrons over this temperature range were not equilibrated to the bath temperature.

When power is dissipated in a small normal metal volume at low temperatures, electrons can be driven far out of equilibrium with the phonon bath, thus leading to elevated electron temperatures \cite{1994PhRvB..49.5942W}. This hot electron effect is characterized by an electron temperature given by $T_e = (P/(\Sigma \Omega)+T_{ph}^5)^{1/5}$ where $P$ is the power dissipated in the shunts, $\Sigma = (1.2\pm 0.4) \times 10^{9} ~\mathrm{Wm^{-3}K^{-5}}$ is the electron-phonon coupling constant for Pd \cite{2007PhRvB..75j4303V}, the phonon temperature $T_{ph} \approx 0.35~\mathrm{K}$ is approximated as the bath temperature, and $\Omega = 12~\mathrm{\mu m}^3$ is the effective volume of our shunts. This volume includes an extra portion of Pd beyond the attachment point to the SQUID and thus does not contribute to the shunt resistance value. The effective volume of this extra Pd is limited by the electron thermal relaxation length that is estimated to be ~$30~\mathrm{\mu m}$ \cite{2009SuScT..22k4007P}. Following this analysis, for a dissipated power of $1~\mathrm{nW}$, typical for the operating point of our MSA, the electron temperature is estimated to be $0.6~\mathrm{K}$. Thus, these hot electron effects likely limit the noise temperature of our MSA as the shunts are significantly hotter than the bath. This also suggests that the electron temperature could be reduced by the addition of more effective metallic cooling volumes for the shunts, thus further improving the noise performance of the MSA.

In conclusion, we have fabricated a microstrip SQUID amplifier that operates at 3.8 GHz with 17 dB of gain and a bandwidth of 150 MHz. The system noise temperature was measured to be 0.55 $\pm~0.13 \mathrm{K}$ at 3.8 GHz and a bath temperature of 0.35 K. The noise temperature is constant for bath temperatures up to 0.55 K suggesting that the noise temperature is limited by electrons above the bath temperature in the shunt resistors, consistent with our estimate of the electron temperature for the power levels dissipated in the SQUID. We estimate the contribution to the system noise from the HEMT to be only a small fraction of the system noise temperature, $\sim 0.12$ K, thus reducing the temperature of the hot electrons in the shunt resistors should have a substantial impact on the system noise temperature. 

The authors acknowledge useful discussions with R. McDermott. This work is supported by the DARPA/MTO QuEST program through a grant from AFOSR. Some of the device fabrication was performed at the Cornell NanoScale Facility, a member of the National Nanotechnology Infrastructure Network, which is supported by the National Science Foundation (Grant ECS-0335765).

%\bibliography{MSA-Tn}

\end{document}